\documentstyle[aps,prl,epsf,preprint]{revtex}
\begin{document}  
\draft

%%%%%%%%%%%%%%%%%%%%%%%%%%%%%%%%%%%%%%%
%     Definitions for style	  %
%%%%%%%%%%%%%%%%%%%%%%%%%%%%%%%%%%%%%%%
\newcommand{\ba}{\begin{eqnarray}}  
\newcommand{\ea}{\end{eqnarray}}  
\newcommand{\M}{{\cal M}}
\renewcommand{\S}{\Sigma}
%%%%%%%%%%%%%%%%%%%%%%%%%%%%%%%%%%%%%%% 
\def\footnote{\def\dummy}  
%%%%%%%%%%%%%%%%%%%%%%%%%%%%%%%%%%%%%%% 

%\begin{titlepage}

\title{ 
	Causality of the brane universe \\
%       \smallskip (\today)      
}
%%%%%%%%%%%%%%%%%%%%%%%%%%%%%%%%%%%%%%% 
\author{
   Hideki Ishihara\footnote{E-mail:ishihara@th.phys.titech.ac.jp}
}
\address{
D\'epartement d'Astrophysique Relativiste et de Cosmologie, \\
Observatoire de Paris, 92195 Meudon, France \\
and\\
        Department of Physics,  
        Tokyo Institute of Technology, \\ 
        Oh-Okayama Meguro-ku, Tokyo 152, Japan
}

\maketitle  

%%%%%%%%%%%%%%%%%%%%%%%%%%%%%%%%%%%%%%% 
\begin{abstract}
Causal structure of the brane universe 
with respect to null geodesics in the bulk spacetime is studied. 
It is pointed out that apparent causality violation is possible 
for the brane universe which contains matter energy. 
It is also shown that there is no \lq horizon problem\rq\ 
in the Friedmann-Robertson-Walker brane universe. 
\end{abstract}
%%%%%%%%% 

%\end{titlepage}

%%%%%%%%%%%%%%%%%%%%%%%%%%%%%%%%%%%%%%% 
\section{Introduction}
%%%%%%%%%%%%%%%%%%%%%%%%%%%%%%%%%%%%%%%			     

Recently, there has been a lot of activity in the idea 
that our universe may be a brane embedded in some 
higher dimensional bulk spacetime. 
If matter fields are confined in the brane while  
gravitational fields live everywhere, 
the extra-dimensions could be large as was thought before\cite{AHD}.
Randall and Sandrum\cite{Randall-Sundrum}(see also\cite{old}) 
have shown the possibility that the brane is embedded in the 5-dimensional 
anti-de Sitter bulk spacetime without compactification 
of the extra-dimension. 
It is remarkable that the four-dimensional standard gravity 
on the brane is recovered with a 
small corrections in the low energy physics
\cite{Randall-Sundrum}\cite{Garriga-Tanaka}
\cite{Shiromizu-Maeda-Sasaki}. 
The self-gravity of the brane plays a important role to single 
out the zero-mode, which corresponds to 4-dimensional gravity, 
as the ground state of the gravitational fields,  
and ${\bf Z}_2$ symmetry, which is suggested from the view point of 
M-theory\cite{Horava-Witten}, 
is required with respect to the brane.

It is also interesting that the existing bulk spacetime gives rise to some 
new physical effects on the brane in the high energy physics, 
where the higher modes would be much excited. 
There are many works concerning the model in various contexts
\cite{cosmology}\cite{others}.

In the Randall-Sandrum model, 
the brane is 4-dimensional intrinsically flat spacetime 
if the tension of the brane is fine tuned so as to cancel 
the bulk cosmological constant. 
However, the brane is curved extrinsically by its self-gravity. 
Then, it would be possible that 
a path appears in the bulk along which an information is carried from 
a point to another on the brane\cite{Chung-Freese}. 
In this case, there would be two possible paths for the information 
propagation: a path confined on the brane and a path in the bulk. 
If the information via the latter path arrives earlier than the former, 
causal structure which is based on the intrinsic geometry of the brane 
should be altered by the existence of the path in the bulk. 
Thus it is important to investigate geodesics along which the 
information propagates in the brane universe model.

%%%%%%%%%%%%%%%%%%%%%%%%%%%%%%%%%%%%%%%%%%%%%%%%%%
\section{Geodesics in the brane universe}
%%%%%%%%%%%%%%%%%%%%%%%%%%%%%%%%%%%%%%%%%%%%%%%%%%

The geodesic is a curve whose tangent vector is parallel propagated
along itself. In the brane universe model, 
there are two kinds of parallel transport: 
one is with respect to the metric $g_{\mu\nu}$ in the bulk, $\M$,  
and the other is with respect to the induced metric $\gamma_{ab}$ 
on the brane, $\S$, which is defined as
\ba
	\gamma_{ab} = g_{\mu\nu} e^\mu_a e^\nu_b,
\ea 
where $e^\mu_a$ is a basis of vectors which span the tangent space of $\S$. 
According to these parallel transports, 
we define two kinds of geodesics which connect two points on $\S$: 
the geodesics on $\S$  whose tangent vectors $u^a$ satisfy 
\ba
	u^a D_a u^c =0 , 
\label{geodesic-S}
\ea	
and the geodesics in $\M$ whose tangent vectors $u^\mu$ satisfy 
\ba
	u^\mu\nabla_\mu u^\nu =0, 
\label{geodesic-M}
\ea
where $D_a$ and $\nabla_\mu$ are the covariant derivatives with respect 
to $\gamma_{ab}$ and $g_{\mu\nu}$, respectively.

For an arbitrary tangent vector field $u^\mu$ on $\Sigma$, 
which is expressed as
\ba
	u^\mu = u^a e^\mu_a, 
%	\gamma_{ab}u^a u^b 	= \pm 1
\ea
these two parallel transports are related by the Gauss-Weingarten equation as
\ba
	u^\mu\nabla_\mu u^\nu 
%		= u^a (\partial_a u^c + \Gamma^c_{ab}u^b )e^\nu_c 
%			- K_{ab}u^a u^b n^\nu \\
		= (u^a D_a u^c)e^\nu_c 
			- K_{ab}u^a u^b n^\nu,
\label{G-W}
\ea
where $K_{ab}$ is the extrinsic curvature of $\S$ and $n^\mu$ is the
normal vector to $\S$. 
It is obvious, as is well known, 
that if $K_{ab}=0$ then every geodesic on $\S$ is also 
a geodesic in $\M$. The hypersurface $\S$ is called \lq totally geodesic\rq\ 
in this case. 
However, the brane universe is not the case.

Self-gravitating brane in the 5-dimensional spacetime has a jump 
in its extrinsic curvatures of 
both sides, say $K^+_{ab}$ and $K^-_{ab}$.  They should satisfy 
the metric junction conditions\cite{Israel},
\ba
	K^+_{ab} -K^-_{ab}= -\kappa^2 ( S_{ab} - \frac13 S \gamma_{ab}),
\label{metric-junction}
\ea
where $\kappa^2$ is the 5-dimensional gravitational constant, 
$S_{ab}$ is the surface energy-momentum tensor of the brane and 
$S$ is its trace. 
Imposing ${\rm\bf Z}_2$ symmetry with respect to $\Sigma$, 
we get 
\ba
	K^+_{ab}= -K^-_{ab}= -\frac12 \kappa^2 ( S_{ab} - \frac13 S \gamma_{ab}). 
\label{Kab-brane}
\ea
The negative sign in front of $K^-_{ab}$ comes from the convention of 
the direction of the normal vector $n^\nu$. 
Hereafter, we consider only + side of the system and drop the symbol +.

For a while, we concentrate ourselves on the vacuum brane 
whose surface energy-momentum tensor is expressed as
\ba
	S_{ab} = -\sigma \gamma_{ab}, 
\label{vac-energy}
\ea
where $\sigma$ is the surface tension of the brane. 
From eqs. (\ref{Kab-brane}) and (\ref{vac-energy}) we get
\ba
	K_{ab}= -\frac16\kappa^2 \sigma \gamma_{ab}. 
\label{Kab-vac-brane}
\ea

For a timelike vector fields $u^a$, it holds that $K_{ab}u^au^b \neq 0$. 
Then, eq.(\ref{G-W}) implies that eq. (\ref{geodesic-M}) 
does not hold for the vector fields on $\S$ satisfying eq.(\ref{geodesic-S}). 
Therefore, a timelike geodesic in $\S$ is not a geodesic in $\M$. 
In other words, an inertial motion on $\S$ is an accelerated motion in $\M$. 

Similarly, a spacelike geodesic in $\S$ is not a geodesic in $\M$. 
In contrast, it is interesting the case of null geodesics. 
Since $K_{ab}k^ak^b = 0$ for null vector fields $k^a$ on $\S$, 
any null geodesic in $\S$ is also a null geodesic in $\M$.

Another property of geodesics is that spacelike (timelike) geodesics 
extremize the proper length (the proper time)
of curves connecting given two points.
We consider whether the geodesics on $\S$ have this property. 

First, we consider the vacuum brane with a positive tension $\sigma>0$. 
$K_{ab}u^au^b < 0$ for a spacelike vector $u^a$ means that 
$\S$ is concave towards $\M$ in the spacelike direction. 
Then for a spacelike geodesic connecting two points on $\S$ 
there are spacelike \lq short cuts\rq\ in $\M$ 
whose length are shorter than the geodesic. 
The shortest curve among the \lq short cuts\rq\ 
is the spacelike geodesic in $\M$. 
On the contrary, $K_{ab}u^au^b  >0$ 
for a timelike vector $u^a$ means that $\S$ is convex towards $\M$ 
in the timelike direction. 
Then a timelike geodesic on $\S$ has still the longest proper time
among curves which connect the endpoints on $\S$. The set of timelike 
geodesics in $\M$ which connect two points on $\S$ is empty 
because the spacetime is singular at the place 
of the self-gravitating brane. 
%As for the null direction, since $\S$ is flat extrinsically 
%a null geodesics on $\S$ is also a null geodesic in $\M$. 

When the tension of the brane is negative, 
there are timelike \lq short cuts\rq\ in $\M$
but there is no spacelike \lq short cut\rq. 
A null geodesic on $\S$ is also a null geodesic in $\M$ again. 
However, since matter fields is confined on $\S$ 
in the brane universe model, 
the timelike geodesics in $\M$ are not of much interest 
because there is no matter particle which moving along 
the curve off the brane.

%%%%%%%%%%%%%%%%%%%%%%%%%%%%%%%%%%%%%%%%%%%%%%%%%%
\section{Causality}      
%%%%%%%%%%%%%%%%%%%%%%%%%%%%%%%%%%%%%%%%%%%%%%%%%%

\begin{figure}[h]
  \begin{center}
    \leavevmode
    \epsfxsize=10cm
    \epsfbox{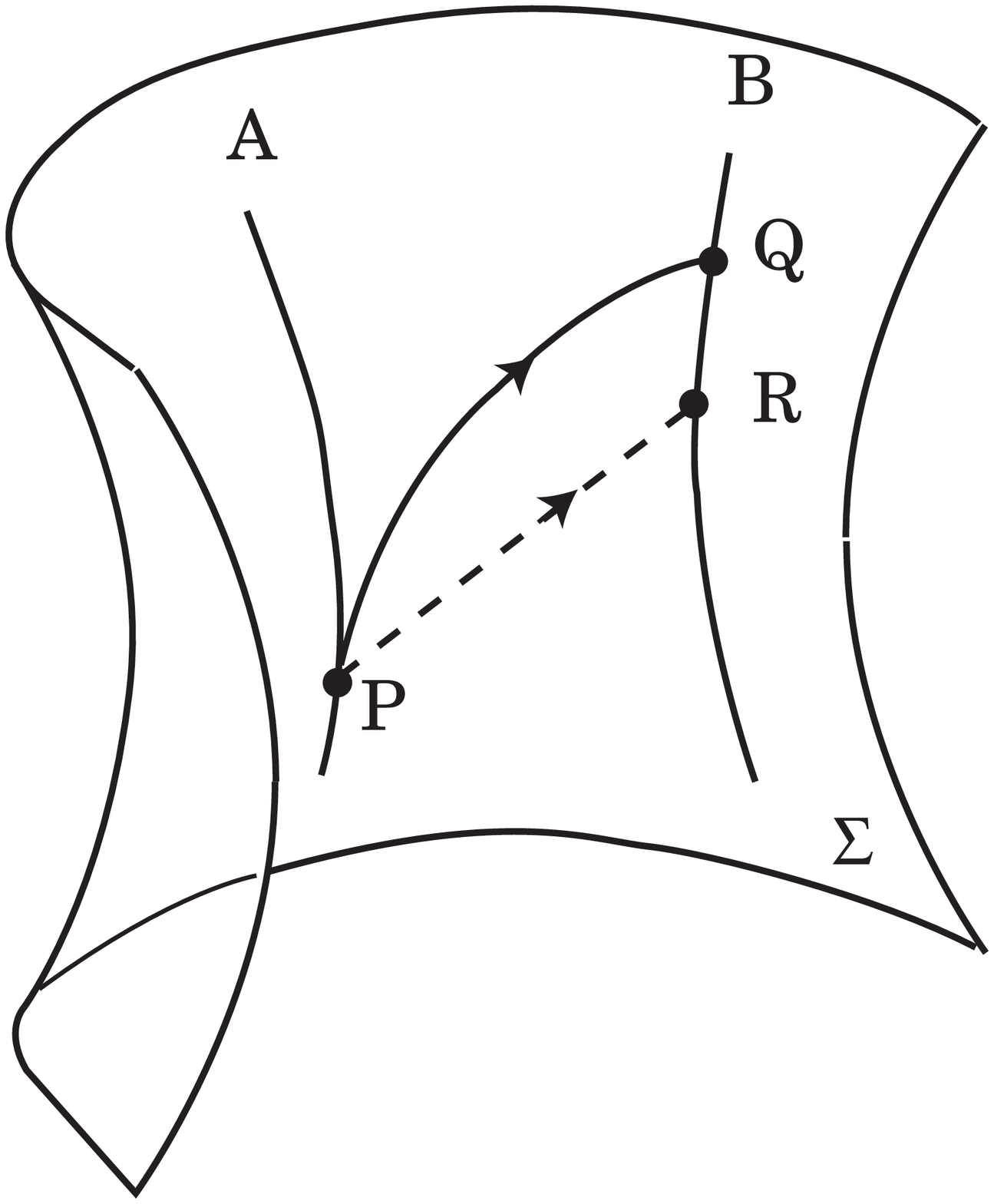}
    \caption{
Possible two paths for an information propagation. 
Solid curve PQ is a null geodesic on $\S$ and broken line PR is a 
null geodesic in $\M$.
}   
    \label{shortcut}
  \end{center}
\end{figure}

Now, let us consider the causality of the brane universe. 
Consider two observers A and B on $\S$. 
The observer A sends a signal at a point P 
towards the observer B by using a massless field on $\S$ 
with the light velocity, 
and the observer B receives the signal at a point Q. 
The points P and Q are connected by a null geodesic on $\S$
(see Fig.\ref{shortcut}).

For the vacuum brane with ${\bf Z}_2$ symmetry, 
there is the unique path for the fastest information propagation 
from an observer to another 
because a null geodesic on $\S$ is also a null geodesic in $\M$. 
However, when we consider a realistic universe which contains 
matter energy on the brane,  
\lq short cuts\rq\ of signal would emerge in $\M$. 
The surface energy-momentum of the brane can take the form 
\ba
	S_{ab} = -\sigma \gamma_{ab} + T_{ab},
\label{EM-matter}
\ea
where $T_{ab}$ is the energy-momentum tensor of the matter on $\S$. 
For ordinary matter it is natural that a energy condition 
$T_{ab}k^ak^b>0$ holds. 
Then we have
\ba
	K_{ab}k^ak^b= -\frac12 \kappa^2 ( S_{ab} - \frac13 S \gamma_{ab})k^ak^b
		= -\frac12 \kappa^2 S_{ab} k^ak^b
		= -\frac12 \kappa^2 T_{ab} k^ak^b < 0. 
\ea
It means that $\S$ is concave towards $\M$ in the null direction.
Then we can find a null geodesic in $\M$ which connects 
P and R on $\S$. 
It is easy to show that the point R exists in the past of the point Q 
defined with respect to the null geodesics on $\S$. 
The points which separate spacelike with respect
to $\gamma_{ab}$ is connected by a causal curve in $\M$. 
That is, the causality in the view point of $\S$ is apparently violated.  
True causality should be defined by the null geodesics in $\M$. 

The magnitude of the apparent causality violation becomes larger 
when the the matter on the brane becomes more dense. 
Then there are two typical 
cases in which the effect becomes important. 
One is gravitational collapse of the matter and the 
other is the early stage of the expanding universe.
In the reset of this paper, we consider the causal structure in 
the early stage of the expanding universe. 

%%%%%%%%%%%%%%%%%%%%%%%%%%%%%%%%%%%%%%%%%%%%%%%%%%%%%%%%%%%%%%%%
\section{Horizon problem } 
%%%%%%%%%%%%%%%%%%%%%%%%%%%%%%%%%%%%%%%%%%%%%%%%%%%%%%%%%%%%%%%%

Let us consider the Friedmann-Robertson-Walker (FRW) brane universe 
embedded in the 5-dimensional anti-de Sitter bulk with a negative 
cosmological constant $\Lambda = -4/l^2$, 
for simplicity.
We assume the energy-momentum tensor of the matter takes the form 
of a perfect fluid, 
\ba
	T_{ab} = (\rho+p) u_a u_b + p \gamma_{ab}, 
\label{perfect-fluid}
\ea
where $\rho$ and $p$ are energy density and pressure 
of the fluid which are separated by the brane tension. 

From the junction condition with ${\bf Z}_2$ symmetry (\ref{Kab-brane}) for 
the surface energy-momentum tensor (\ref{EM-matter}) and (\ref{perfect-fluid}) 
we get the modified Friedmann equation\cite{cosmology}
\ba
	\left( \frac{1}{a}\frac{d a}{d\tau}\right)^2 
		= \frac{8\pi}{3} \frac{\kappa^4\sigma}{6} \rho - \frac{k}{a^2} 
			+ \frac{\kappa^4}{36}\rho^2,
\label{Hubble-eq}
\ea 
where fine tuning $ 1/l^2 = \kappa^4 \sigma^2/36$ has been done 
so as to eliminate the 4-dimensional effective cosmological constant 
on $\S$. 
Here, $a(\tau)$ is the scale factor of the FRW brane universe, 
$\tau$ is the proper time of the co-moving matter on the brane 
and $k=+1,0,-1$ for the closed, flat and open universe model, respectively. 
The last term in the right hand side is a new term for 
the brane universe model. 
The energy conservation law has a usual form
\ba
	\frac{d}{d\tau}(\rho a^3) + p \frac{d}{d\tau}(a^3) =0. 
\label{cons-law}
\ea

We consider the closed FRW universe case, in this paper, 
because it is the easiest case to see the embedding. 
The essential feature which we will see below is same for 
flat and open FRW universe cases as will be discussed 
in the separated paper\cite{Ishihara}. 

We use the static coordinate for the 5-dimensional 
anti-de Sitter bulk spacetime, $\M$,
\ba
	ds_5^2 = -\left(1+\frac{r^2}{l^2}\right)dt^2 
		+\left(1+\frac{r^2}{l^2} \right)^{-1} dr^2 
		+ r^2 d\Omega_3^2,
\label{ads}
\ea
where  $d\Omega_3^2$ is the metric of unit $S^3$. 
The embedding of the FRW brane, $\S$, in $\M$  
is given by
\ba
	r = a(\tau), \quad t = T(\tau), 
\label{embed}
\ea
where $a(\tau)$ is a solution of eq.(\ref{Hubble-eq}) and $T(\tau)$ is 
determined by
%%%%%%
\ba
	-\left(1+\frac{a^2}{l^2}\right)\left(\frac{dT}{d\tau}\right)^2 
   + \left(1+\frac{a^2}{l^2} \right)^{-1}\left(\frac{da}{d\tau}\right)^2 = -1.
\label{normalization}
\ea
%%%%%%

\begin{figure}[h]
  \begin{center}
    \leavevmode
    \epsfxsize=12cm
    \epsfbox{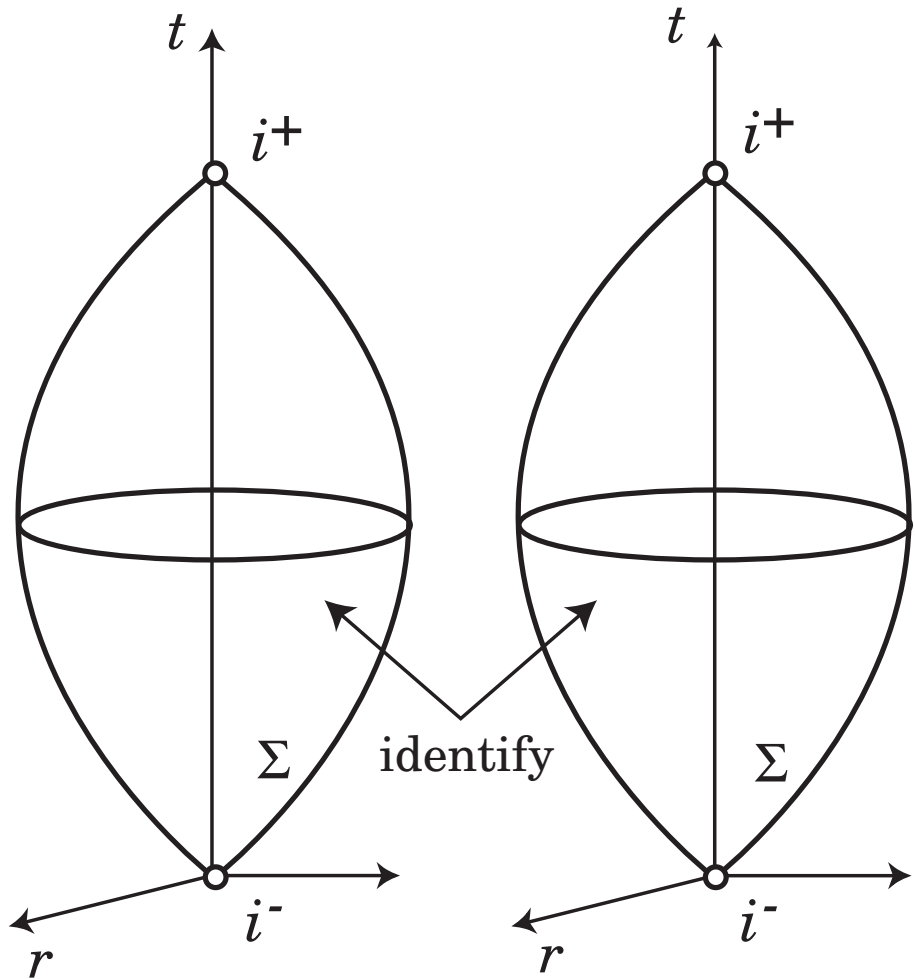}
    \caption{
Embedding of the closed FRW brane universe. 
The bulk is surrounded by the brane, where 
$S^3$ of the FRW brane universe is reduced to $S^1$ in this figure. 
$i^-$ and $i^+$ denote the initial and final singularities. 
The total spacetime is obtained by identification of the surfaces $\S$ 
of two copies.
}   
    \label{embedding}
  \end{center}
\end{figure}

It is easy to see the qualitative behavior of the scale factor $a(\tau)$ 
from eqs.(\ref{Hubble-eq}) and (\ref{cons-law}).
For the equation of state 
$p=\beta \rho$ with a constant $\beta > -2/3$, 
the closed universe begins at the initial singularity, expands 
to the maximum radius, and recollapse to the final singularity in a 
time symmetric manner. (see Fig.\ref{embedding}.)
The initial behavior of the scale factor is determined by the last term 
in the right hand side of eq.(\ref{Hubble-eq}) 
as $a \propto \tau^{1/(3(\beta+1))}$.
From eq.(\ref{normalization}) it is obtained that the 
asymptotic behavior of $a$ at the initial singularity is $a = t$.
Therefore, $\S$ is a timelike surface  
which is tangent to the 
light cone at the initial and final singularities as shown 
in Fig.\ref{embedding}.

\begin{figure}[h]
  \begin{center}
    \leavevmode
    \epsfxsize=9cm
    \epsfbox{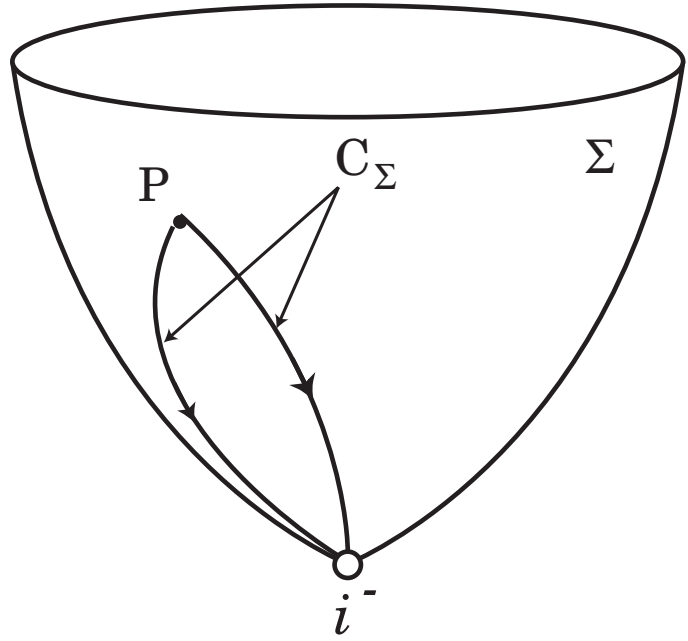}
    \caption{
Past directed null geodesics on the brane near the initial singularity. 
}   
    \label{internal}
  \end{center}
\end{figure}

To see the causal structure near the initial singularity, 
take a arbitrary point P in the vicinity of it.
Null geodesics on $\S$, which are obtained in a usual manner, 
starting from the point P 
converge backward to the initial singularity as the usual 
FRW cosmology (see Fig.\ref{internal}).
The set of past directed null geodesics on $\S$ from P, 
say $\mbox{C}_\S$, is the boundary of causal past of P 
defined with respect to the null geodesics on $\S$.  
If we restrict ourselves to consider physical processes on $\S$, 
the observer at P can communicate 
a small part of observers starting from the initial singularity 
whose world lines intersect $\mbox{C}_\S$. 
That is, the initial singularity is spacelike 
as is in the usual FRW universe model. 

\begin{figure}[h]
  \begin{center}
    \leavevmode
    \epsfxsize=10cm
    \epsfbox{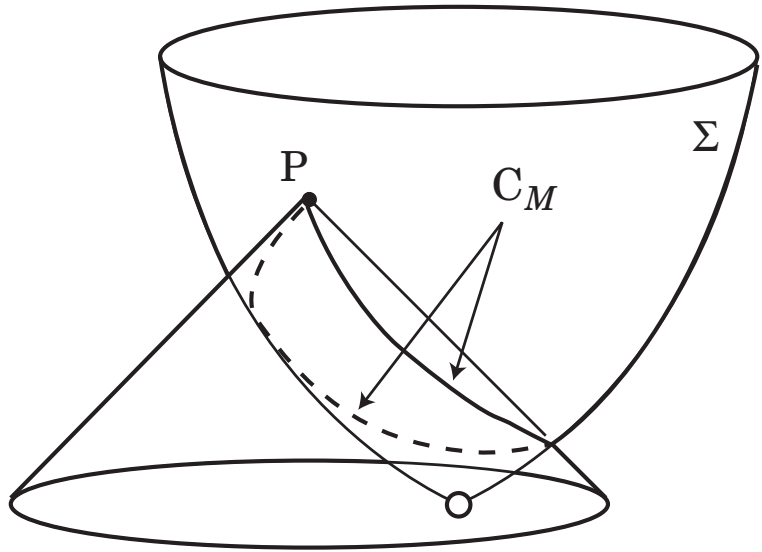}
    \caption{
Past directed light cone near the initial singularity. 
}   
    \label{external}
  \end{center}
\end{figure}

On the other hand, if we take null geodesics in $\M$ into account, 
the causal structure becomes different. 
The light cone of P, which is the subset of $\M$ generated 
by null geodesics in $\M$ starting from P on $\S$, 
is shown in Fig.\ref{external}. 
%%%%%%
If P is taken so close to the initial singularity that 
$r^2/l^2$ in the metric (\ref{ads}) can be neglected , the light
cone illustrated in this figure is approximately same as 
in Minkowski spacetime. 
%%%%%%
Let $\mbox{C}_\M$ be an intersection of the light cone of P 
and the brane universe $\S$. $\mbox{C}_\M$ is the set of points on $\S$ with 
which the observer at P can communicate. 
Since $\mbox{C}_\M$ is a spacelike surface on $\S$ 
which intersects all timelike world lines starting 
from the initial singularity, 
it is seen that the observer at P can communicate 
all observers in the universe.
In contrast to the usual FRW universe, 
there is no particle horizon in the brane universe. 
The initial singularity is pointlike from the view point 
of the null geodesic in $\M$.

By the qualitative analysis in this paper, it is shown that 
the horizon problem in the FRW brane universe does not exist 
since gravitons in the excited modes (KK-modes), 
which travel along null geodesics in the bulk, can carry an information 
from a point on the brane universe to another quickly. 
However, it is still open question 
whether the enough amount of gravitons in exited modes
could homogenize the early universe or not. 
As noted before, even an inertial particle on $\S$ is an accelerated 
particle in $\M$. The highly accelerated particles in the early 
universe emit and absorb gravitational waves which propagate in the bulk. 
It would be envisaged that the backreaction of these processes would 
cause the homogenization and isotropization of the brane universe. 
The closed FRW brane universe is a \lq spherically symmetric\rq\ 
solution in the 5-dimensional sense. 
By the 5-dimensional version of Birkoff's theorem, it emits 
no more gravitational wave. 
Then, the isotropic and homogeneous brane universe would be a final state 
of the expanding universe.  

%%%%%%%%%%%%
Finally, it should be noted that when the energy density 
of the matter reaches order of the brane tension, 
the brane description, which is assumed in this paper, 
would not be valid. It is an interesting future problem 
to obtain a complete picture of the 
causality of the brane universe including quantum corrections. 
%%%%%%%%%%%%

%%%%%%%%%%%%%%%%%%%%%%%%%%%%%%%%%%%%%%%
\begin{acknowledgements}
The author would like to thank N. Deruelle for her encouragement. 
He gratefully acknowledges the hospitality of DARC, Paris Observatory, 
Meudon. 
\end{acknowledgements}
%%%%%%%%%%%%%%%%%%%%%%%%%%%%%%%%%%%%%%%

%%%%%%%%%%%%%%%%%%%%%%%%%%%%%%%%%%%%%%%

%%%%%%%%%%%%%%%%%%%%%%%%%%%%%%%%%%%%%%%%%%%%%%%%%%%%%%%%%%%%%%%%%%%%%%%%%

%

\newpage

%%%%%%%%%%%%%%%%%%%%%%%%%%%%%%%%%%%%%%%%%%%%%%%%%%%%%%%%%%%%%%%%%%%%%%%%%

\end{document}